\documentclass[twocolumn,prl,showpacs,floatfix]{revtex4}
\usepackage{graphicx}
\usepackage{epsfig}
\usepackage{rotating}
\usepackage{amsmath}
\usepackage{amsfonts}
\usepackage{amssymb}
\usepackage{enumerate}
\usepackage{longtable}
\usepackage{dcolumn}
\usepackage{bm}

\begin{document}

\title{High temperature series expansion study of the Heisenberg antiferromagnet on the hyperkagome lattice: Comparison with Na$_4$Ir$_3$O$_8$.
}

\author{R. R. P. Singh}
\affiliation{Department of Physics, University of California Davis, CA 95616, USA}

\author{J. Oitmaa }
\affiliation{School of Physics, The University of New South Wales,
Sydney 2052, Australia}

\date{\rm\today}

\begin{abstract}
We develop high temperature series expansions for 
$\ln{Z}$ and the uniform structure factor of the spin-half Heisenberg
model on the hyperkagome lattice to order $\beta^{16}$. These
expansions are used to calculate the uniform susceptibility ($\chi$), 
the entropy ($S$),
and the heat capacity ($C$) of the model
as a function of temperature. Series
extrapolations of the expansions converge well down to a temperature of 
approximately $J/4$. 
A comparison with the experimental data for Na$_4$Ir$_3$O$_8$
shows that its magnetic susceptibility
is reasonably well described by the model with
an exchange constant $J\approx 300 K$, but there are also additional smaller
terms present in the system. The specific heat of the model has two
peaks. The lower temperature peak, which is just below our range of
convergence contains about 40 percent of the total entropy. Despite
being a 3-dimensional lattice, this model shares many features with
the kagome lattice Heisenberg model and the material 
must be considered a strong candidate for a quantum spin-liquid.

\end{abstract}


\maketitle

\section{Introduction}

Despite many experimental
and theoretical studies, the 
spin-half kagome lattice
Heisenberg Model remains incompletely understood and a subject of
intense ongoing research.\cite{fukuyama,helton,mendels}
The kagome lattice is a two-dimensional
lattice of corner sharing triangles, where classical antiferromagnetic models
lead to exponential ground state degeneracy. Whether quantum
fluctuations lift this degeneracy and lead to an ordered state, or
select a valence-bond order, or lead to a 
resonating valence-bond quantum spin-liquid possibly
with topological order and fractional excitations remains a topic
of great interest. Numerical evidence shows that many distinct singlet states
are very close to each other in energy.\cite{yan,ran,iqbal,poilblanc,oleg,singh-huse,khatami,lecheminant,sindzingre,mila}

The hyperkagome lattice
is a three-dimensional lattice of corner sharing triangles, where
similar ground state degeneracy at the classical level and possible
competing phases with nearly the same energy in presence of
strong quantum fluctuations remains valid. Experimental
discovery\cite{okamoto} of the material  Na$_4$Ir$_3$O$_8$, where the $Ir$ ions
form a hyperkagome lattice, with strong signatures of
a gapless quantum spin-liquid phase has led to much excitement.
Theoretical proposals for this system also include
various quantum spin-liquids as well as 
valence bond order.\cite{lawler1,lawler2,bergholtz,zhou}

In this paper, we develop high temperature series expansions for
$\ln{Z}$ and the uniform structure factor of the spin-half 
hyperkagome lattice Heisenberg model to order $\beta^{16}$.
These series expansions are used to calculate the uniform
susceptibility, the entropy and the heat capacity of the model.
Pad\'e and d-log Pad\'e approximants show that series extrapolations
converge well down to a temperature of approximately $J/4$.
At this temperature $40$ percent of the total entropy of the
system is still present. The heat capacity, which has a short-range
order peak around $T=2J/3$ shows hints of another rise below $T=J/4$,
leading to a sharp peak in $C/T$. These results are reminiscent of
similar results on the kagome lattice.\cite{elser,elstner,misguich,rigol}

Although the susceptibility data
can be well fit by a Curie-Weiss law over a wide temperature range,
the effective Curie-Weiss constant keeps changing with the
region of fit.\cite{coldea}
Comparison with the experimental susceptibility data
shows that the material is described by
a $J$ value of approximately $300K$.\cite{lawler1,lawler2}
This $J$ value is consistent with more recent high-field 
measurements.\cite{okamoto-pc}
There are also clear deviations
below $T=J/2$, where the series extrapolation are still well convergent.
These presumably arise from smaller exchange terms induced by spin-orbit
couplings, such as the Dzyaloshinski-Moria interactions, which are
not included in our theoretical studies.\cite{micklitz,chen-balents}
Similar anisotropy terms are also known to be present in the
kagome antiferromagnets.\cite{mendels,rigol,cepas}

Experimental determination
of the magnetic heat capacity requires subtraction of non-magnetic
contributions, which becomes increasingly uncertain with increasing
temperature. Our study can not
shed light on the low temperature power-laws observed in the
material. At temperatures above $J/3$ which translates
to $\approx 100 K$ for the material, where our results are
well converged, there is a clear discrepancy between the
experiments and the model calculations. We show that these discrepancies
can be reconciled 
if one assumes some additional harmonic modes below $300K$
that have not been fully eliminated by the subtraction
procedure in the experimental paper.\cite{okamoto}

\section{Model and Series Expansions}
We study the nearest-neighbor spin-half Heisenberg model
\begin{equation}
{\cal H}=J\sum_{i,j} \vec S_i \cdot \vec S_j.
\end{equation}
Here, the sum runs over each nearest-neighbor pair of spins
on the hyperkagome lattice. We develop high temperature
series expansions for $\ln{Z}$ and the uniform structure factor,
which is also $T\chi$, where $\chi$ is the uniform susceptibility.
These quantities are given by
\begin{equation}
\ln{Z}=\ln{Tr \exp{-\beta {\cal H}}},
\end{equation}
and
\begin{equation}
T\chi=\sum_{i,j} <S_i^z S_j^z>.
\end{equation}
Here the sum runs over all the spins of the lattice,
and the angular brackets denote a thermal expectation value.
Details of the method for calculating the series expansions
can be found in the literature.\cite{book}
Let us write
\begin{equation}
{\ln{Z}\over N}=\sum_n a_n ({-\beta J\over 4})^n,
\end{equation}
where $N$ is number of sites, and,
\begin{equation}
{T\chi\over N}=\sum_n b_n ({-\beta J\over 4})^n.
\end{equation}
The coefficients $a_n$ and $b_n$ upto $n=16$
are given in Table 1. Note that the series coefficients are
valid for both the ferromagnetic and the antiferromagnetic couplings.
The factors of $(-1)^n$ are needed for the antiferromagnetic model,
where $J>0$.

\begin{table}
\begin{tabular}{|l|c|c|}
\hline
$n$ & $a_n$ & $b_n$ \\ \hline
0 & $\ln{2}$ & 0.25 \\
1 & 0.000000000000D+00  &  0.100000000000D+01\\
2 & 0.300000000000D+01  & 0.200000000000D+01 \\
3 & 0.000000000000D+00  &  0.100000000000D+01 \\
4 & -0.850000000000D+01 &  -0.666666666667D+00\\ 
5 & 0.000000000000D+00  & 0.134666666667D+02\\
6 & 0.465333333333D+02  & 0.352888888889D+02\\
7 & 0.213333333333D+01  & -0.699206349206D+02\\
8 & -0.312973809524D+03 & -0.237619047619D+03\\
9 & -0.354539682540D+02 &  0.745729100529D+03\\
10 & 0.235000994709D+04 & 0.236886624339D+04\\
11 & 0.467922116402D+03 & -0.685090572391D+04\\
12 & -0.188903919384D+05 &  -0.231225320571D+05\\
13 & -0.570519634440D+04 & 0.656040985160D+05\\
14 & 0.158860661564D+06  &  0.236506418283D+06\\
15 & 0.668440957641D+05  & -0.623070199393D+06\\
16 & -0.137796110009D+07 & -0.244735886559D+07\\ \hline
\end{tabular}
\caption{\label{series} Series expansion coefficients for $\ln{Z}$
and uniform structure factor for the hyperkagome lattice Heisenberg
model.}
\end{table}

\section{Series Analysis and Results}
\begin{figure}
\begin{center}
 \includegraphics[width=7cm,angle=270]{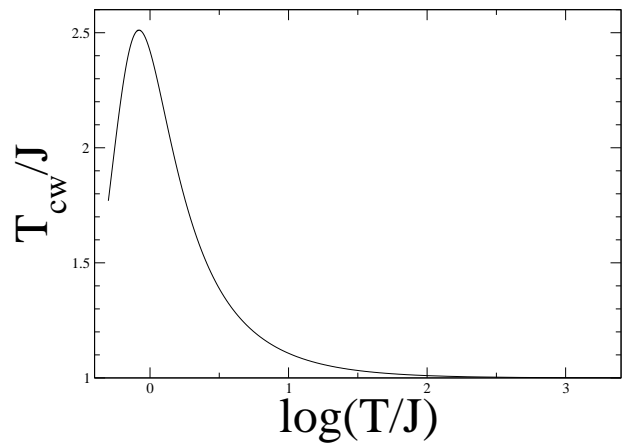}
\caption{\label{fig1} 
Variation of the effective Curie-Weiss constant with temperature
for the Heisenberg Antiferromagnet on the hyperkagome lattice.
}
\end{center}
\end{figure}
We begin with an analysis of the uniform susceptibility series.
To convert the susceptibility into emu/mole, we need to
multiply $T\chi$ by a factor of 
$N_A g^2 \mu_B^2/4k_B=0.0938 g^2$. At high temperatures the
susceptibility can be fit to a Curie-Weiss law, whose asymptotic value for our
model is $T_{cw}^{\infty}=J$. However, as discussed by Zheng et al,\cite{coldea}
the effective Curie-Weiss parameter obtained by fitting the
inverse susceptibility versus temperature data to a straight line
slowly changes with temperature. It can be defined around a temperature $T$ as
\begin{equation}
T_{cw}(T)=-T -{\chi\over d\chi/dT}.
\end{equation}
A plot of this effective Curie-Weiss constant as a function of $\log(T/J)$
is shown in Fig.~1. We see that even at temperatures of $10 J$ the
effective Curie-Weiss constant is $10$ percent off its asymptotic value.
At temperatures around and below $J$ the effective Curie-Weiss constant
is larger than $2J$. This means that the exchange constant must
be a factor of 2 or more smaller than determined by Okamoto 
et al.\cite{okamoto} Indeed, $J=300 K$ and $g=1.99$ leads to an excellent
fit of the experimental data above $T=150 K$. Since the $g$-factor is
not independently known, a finer fit is not useful. We note that
the series extrapolation converges well down to about $J/4$ and below
$J/2$, there are clear deviations between the experimental data and the
Heisenberg model. We attribute these to spin-orbit couplings, which
must be present in the material.\cite{chen-balents}
Nevertheless, the susceptibility comparison
shows that the nearest-neighbor Heisenberg model is a good
starting model for the material.\cite{micklitz} All other
terms are presumably an order of magnitude smaller.

\begin{figure}
\begin{center}
 \includegraphics[width=7cm,angle=270]{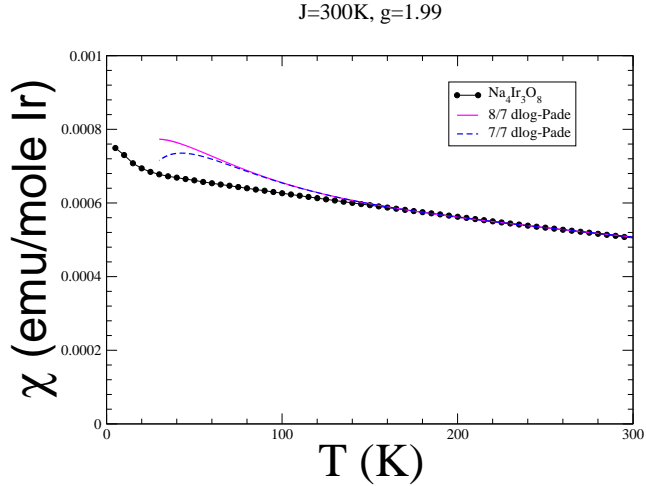}
\caption{\label{fig2} 
A comparison of the uniform susceptibility data of Na$_4$Ir$_3$O$_8$
versus temperature with the Heisenberg model for $J=300K$
and $g=1.99$.
}
\end{center}
\end{figure}

For the rest of the study, we will fix the $J$ value to $300 K$.
Series extrapolations for the heat capacity are shown in Fig.~3.
There is a peak associated with short-range order around a temperature
of $200 K$. This peak is well captured in our calculation. However,
the extrapolations do not converge well below the peak. 
Experiments show a $T^2$ heat capacity at low temperatures.
If we use an [N+2/N] pade to represent the heat capcity, it
ensures a $T^2$ dependence. All such extrapolations are close
to the [6/4] approximant shown. However, they can not be right. The integration
of the heat capacity implies a large missing entropy of
order $40$ percent. A similar problem was observed in extrapolating
the high temperature series for the kagome lattice 
Heisenberg model.\cite{elstner} If instead we use d-log Pad\'e
approximants to extrapolate the series, we find that the
majority of the approximants show a second rise in the
heat capacity at a temperature below $100 K$. However,
once again the convergence breaks down as the second peak arises.
The method developed by Misguich and Bernu\cite{misguich} may
be helpful in this regard. However, their method is based on taking advantage
of our knowledge of the ground state and the low temperature properties in the
extrapolation of the high temperature expansions. 
Since we know even less about the ground
state properties of the hyperkagome lattice than we do about the
kagome lattice, it is unclear how useful it would be at this stage.

\begin{figure}
\begin{center}
 \includegraphics[width=7cm,angle=270]{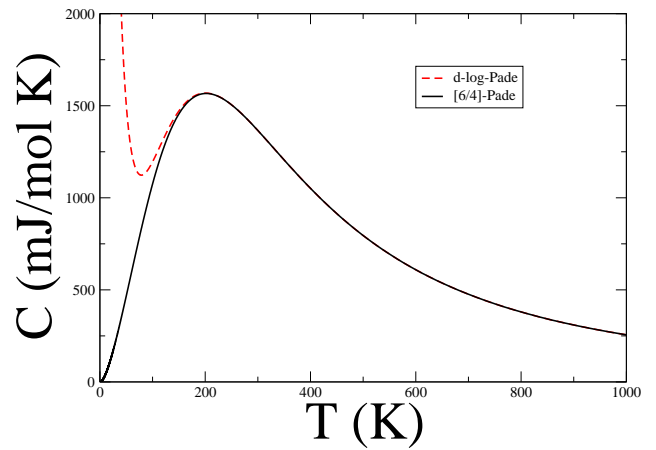}
\caption{\label{fig3} 
The molar heat capacity for the hyperkagome lattice Heisenberg model with $J=300 K$.
The [N+2/N] pade approximant ensures a $T^2$ behavior at low temperatures.
But, it has a $40$ percent missing entropy. The d-log Pad\'e
approximants imply a second peak below $T=100 K$.
}
\end{center}
\end{figure}

\begin{figure}
\begin{center}
 \includegraphics[width=7cm,angle=270]{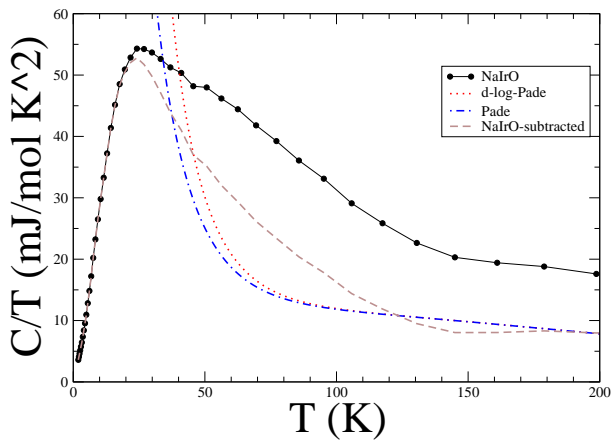}
\caption{\label{fig4} 
Comparison of $C/T$ between theory and experiments. The line with 
filled circles is the experimental data from Ref.\onlinecite{okamoto}.
The solid lines show representative behavior from series
extrapolations. The dashed line is experimental data from which
additional subtractions corresponding to a heat capacity of $0.25$ moles of
phonons at a single frequency ($200K$) has been carried out.
}
\end{center}
\end{figure}

\begin{figure}
\begin{center}
 \includegraphics[width=7cm,angle=270]{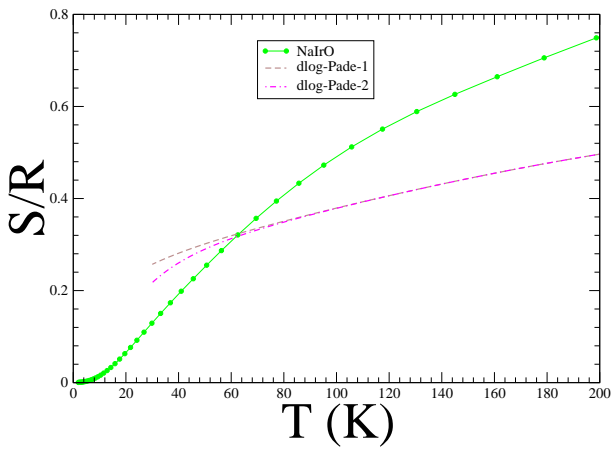}
\caption{\label{fig5} 
The entropy function calculated for the hyperkagome lattice
Heisenberg model and the experimental material.
}
\end{center}
\end{figure}

A comparison of the experimental $C/T$ data with the theory is shown in
Fig.~4. We clearly cannot address the low temperature power-laws, 
which occur at temperatures well below our range of convergence.
The discrepancy at higher temperatures can possibly be attributed to
an incomplete subtraction. Experimental determination of the
magnetic specific heat requires subtraction of phonon and other
contributions. When the $J$ values are comparable to room temperature,
this becomes difficult as the phonon contributions are not simply
power-laws. For this material, the non-magnetic Na$_4$Sn$_3$O$_8$
was used to subtract the phonons. But such a subtraction need not
be accurate above some temperature. In fact, as shown in Fig.~5, the
entropy obtained from the experimental data exceeds the theoretical
curves above 70 K and exceeds $\ln{2}$ before $200$ K. In our calculations, the
entropy is well converged down to about $70$ K. It is also
much less sensitive upto these temperatures to extrapolation methods,
and to addition of other smaller exchange constants. 

We believe, there can be two
reasons for this discrepancy. First, the subtraction is incomplete.
Indeed, if we further subtract harmonic oscillator modes with a density of 
states of about $0.25$ moles at a single frequency of $200K$ that can
bring the data into good agreement with experiments,
as shown in Fig.~4. A second
possibility is that the materials have additional entropy associated
with itinerant degrees of freedom because they are not too far
from a metal insulator transition. This issue can presumably
be addressed by further experiments.

\section{Conclusions}
In this paper, we have developed high temperature series expansions for 
various thermodynamic properties of the Heisenberg model on the
hyperkagome lattice. These expansions enable us to compute various
thermodynamic properties, such as the uniform susceptibility, the
heat capacity and the entropy down to a temperature of $J/4$. Comparison with
the experimental data for Na$_4$Ir$_3$O$_8$ suggests that the
Heisenberg model is a good starting model for the material
with an exchange constant $J\approx 300 K$. However, we find that there
must be additional terms also present in the material. 
The observed low temperature power-laws can not be addressed
by our calculations. We have suggested possible resolutions
to discrepancies in higher temperature
heat capacity and entropy data for the material.

\begin{acknowledgements}
We would like to thank Dr. Y. Okamoto and Dr. H. Takagi
for many fruitful communications and for sending us their 
experimental data.
We are grateful for the computing resources provided by the 
Australian Partnership for Advanced Computing (APAC) National Facility.
This work is supported in part by NSF grant number  DMR-1004231.
\end{acknowledgements}


\end{document}